\documentstyle[sprocl]{article}

\def\be{\begin{equation}}
\def\ee{\end{equation}}
\def\bea{\begin{eqnarray}}
\def\eea{\end{eqnarray}}

\newcommand{\bkappa}{\mbox{\boldmath $\kappa$}}

\begin{document}

\title{ANATOMY OF THE PROTON STRUCTURE FUNCTIONS
IN $\kappa$-FACTORIZATION}

\author{I.P. IVANOV, N.N. NIKOLAEV}

\address{IKP (Theorie) Forschungszentrum J\"ulich, Germany\\
E-mails: i.ivanov@fz-juelich.de, n.nikolaev@fz-juelich.de}

\maketitle
\abstracts{We present the first experiment-based
parameterization of differential gluon structure function,
which is called upon in many applications.
We compare $\kappa$-factorization and DGLAP approaches and
analyze properties of DGSF, with special emphasis on
soft-to-hard/hard-to-soft diffusion.}

It has become a common wisdom to present a DIS off
a proton as if occurring off distinct partons (gluons in
small-$x$ domain) in a quantum mechanical probabilistic fashion
(the familiar DGLAP approach), which involves
the integral flux of gluons with transverse
momenta less or equal to $Q^2$ $G(x,Q^2)$,
whose parameterizations are available to the community.
This collinear DGLAP approximation has been suspected
for a long time to break in the small-$x$ domain,
where the amplitude clearly develops the non-DGLAP features.
In this regime one should turn to the original,
diagrammatically straightforward relation
(the so-called $\kappa$-factorization approach):
$$
{d\sigma_{\gamma^{(*)}p}(x,Q^2) \over d\log x} =
\int\limits_0^{{Q^2\over 4x}} {d\bkappa^2 \over \bkappa^2}
\sigma_{\gamma^{(*)}p}(\bkappa^2)
{\cal F}(x,\bkappa^2)\,;\quad
{\cal F}(x,\bkappa^2) =
{\partial G(x,\bkappa^2)\over \partial \log\bkappa^2}
\nonumber
$$
with ${\cal F}(x,\bkappa^2)$
being the Differential Gluon Structure Function (DGSF) ---
the object of the small-$x$ evolution BFKL equation.

In this work we pursued three goals: (1) construct
a practical parameterization of DGSF consistent with
experimental data on $F_{2p}$ both in soft and hard region;
(2) compare physical observables in the DGLAP and
$\kappa$-factorization approaches and quantify the above
statement on restricted applicability of DGLAP approach;
(3) investigate so constructed DGSF, with emphasis on
soft-to-hard/hard-to-soft diffusion.

Constructing an Ansatz for DGSF, we followed a pragmatic strategy:
for hard gluons we made as much use as possible of the existing
parameterizations of $G(x,Q^2)$, which we interpolated into the
soft region in compliance with color singlet constraints, and we
added the purely non-perturbative soft component, which was
interpolated into the hard domain.  This Ansatz has several free
parameters, which we adjusted by matching the DGSF-based
description of $F_{2p}$ with experimental data throughout the
whole $Q^2$ region ($\sigma_{\gamma p}$ in the $Q^2=0$ limit). The
explicit form of parameterization as well as sets of parameters
are given in \cite{main}. Thus, we can state that we extracted
DGSF from experiment and put it into the form of handy
parameterizations, available to the community.

Upon constructing the explicit for of DGSF, we can compare
the integrated gluon density $G_D=\int_0^{Q^2}
{\cal F}(x,\bkappa^2) d\log \bkappa^2$ with the DGLAP output
$G_{pt}$. This comparison reveals that though $G_D$ and
$G_{pt}$ do converge at ultimately high $Q^2$,
they differ significantly at moderate $Q^2$,
divergence being more prominent at smaller $x$.
At $x=10^{-5}$, $G_{pt}$ is two-three times higher than
$G_{D}$ for $Q^2$ up to 100 GeV$^2$. This should be regarded
as a warning against unwarranted applications of
DGLAP-based quantities in the small-$x$ domain.

One of the principal features of the $\kappa$-factorization
approach is the soft-to-hard and hard-to-soft diffusion
phenomenon, which originates from lifting of strong ordering
of transverse momenta, inherent to DGLAP.
The explicit subdivision of our DGSF Ansatz into soft and hard
parts makes it possible to track down effects of soft
gluons in the hard region and vice versa.
We observed that the soft, non-perturbative component
does not vanish and even rises with $Q^2$ in $F_{2p}$,
being a dominant feature of $F_{2p}$ for $Q^2 \le 10$ GeV$^2$
at $x=0.8\cdot 10^{-3}$.

The reverse side of the same phenomenon, the hard-to-soft
diffusion, stands behind the energy rise of the real
photoabsorption cross section. We emphasize that the correct
energy rise was modelled in the approach with manifestly
energy independent soft component: the rise is entirely due
to hard-to-soft diffusion (contrary to the Donnachie-Landshoff
type parameterizations of color dipole cross sections,
where the soft intercept is the input quantity).

Another finding came from small-$x$ growth of different
observables (and their hard parts), which we presented in terms
of effective intercepts. We observed transition from
a strongly $Q^2$-dependent hard intercept in the case
of ${\cal F}(x,\bkappa^2)$ to almost $Q^2$-independent
intercept of the integral quantity $F_{2p}$.
From the Regge theory point of view, this dramatic flattening
leads to a very simple two-pole picture of $F_{2p}$:
almost fixed pole with $\Delta_{hard}\approx 0.4$ hard part
plus energy independent soft constribution.

\end{document}